\documentclass[aps,prd,twocolumn,superscriptaddress,amsfont,amssymb,amsmath,nofootinbib,showpacs,balancelastpage]{revtex4-1}

\usepackage{graphicx,longtable,natbib,mathrsfs,xcolor,ulem,array,float}
\usepackage{txfonts,aas_macros}

\newcommand{\bs}{\boldsymbol}

\newcommand{\T}{\mathcal{T}}
\newcommand{\Msun}{M_\odot}

\begin{document}

\addtolength{\hoffset}{-0.525cm}
\addtolength{\textwidth}{1.05cm}
\title{Spin conservation of cosmic filaments}

\author{Ming-Jie Sheng}
\affiliation{Department of Astronomy, Xiamen University, Xiamen, Fujian 361005, China}

\author{Sijia Li}
\affiliation{Department of Astronomy, Xiamen University, Xiamen, Fujian 361005, China}

\author{Hao-Ran Yu}\email{haoran@xmu.edu.cn}
\affiliation{Department of Astronomy, Xiamen University, Xiamen, Fujian 361005, China}

\author{Wei Wang}
\affiliation{Purple Mountain Observatory, Nanjing, Jiangsu 210034, China}
\affiliation{School of Astronomy and Space Science, 
University of Science and Technology of China,
Hefei, Anhui 230026, China}

\author{Peng Wang}
\affiliation{Leibniz-Institut f\"ur Astrophysik Potsdam, 
An der Sternwarte 16, D-14482 Potsdam, Germany}

\author{Xi Kang}
\affiliation{Purple Mountain Observatory, Nanjing, Jiangsu 210034, China}
\affiliation{Zhejiang University-Purple Mountain Observatory Joint Research Center for Astronomy, 
Zhejiang University, Hangzhou, Zhejiang 310027, China}

\date{\today}

\begin{abstract}
Cosmic filaments are the largest collapsing structure in the Universe.
Recently both observations and simulations inferred that cosmic
filaments have coherent angular momenta (spins).
Here we use filament finders to identify the filamentary structures
in cosmological simulations and study their physical origins,
which are well described by the primordial tidal torque
of their Lagrangian counterpart regions—protofilaments.
This initial angular momenta statistically preserve their
directions to low redshifts.
We further show that a spin reconstruction method can predict the
spins of filaments and potentially relate their spins to the initial
conditions of the Universe.
This correlation provides a new way of constraining and obtaining
additional information of the initial perturbations of the Universe.
\end{abstract}
 

\maketitle

\section{Introduction}\label{sec.intro}
The large scale structure (LSS) of the Universe contains plenty of cosmological information
and enables us to answer questions about the initial state of the Universe \citep{1969ApJ...155..393P}.
The key procedure is looking for linear mappings from the observables at low redshifts to
the properties of the initial perturbations at high redshifts
\citep{2005MNRAS.360L..82R,2021JCAP...06..024M}.
The nonlinear clustering of LSS leaves linear Fourier modes 
only $k\lesssim 0.2\,h\,{\rm Mpc}^{-1}$, and even with reconstruction methods 
\citep{2017PhRvD..95d3501Y,2017MNRAS.469.1968P} the available linear Fourier modes are still limited.
It is thus valuable to find observables that relate to the initial perturbations.

Beside using the locations and velocities of galaxies to study LSS,
the rotations of galaxies provide another degree of freedom to constrain 
the initial conditions and cosmological parameters.
At low redshifts, the galaxy angular momenta (spins) are 
observable via their ellipticity, projection angles,
spiral parities, and Doppler effects \citep{2019ApJ...886..133I}
and are physically related to initial perturbations.
The tidal torque theory explains how the angular momentum of a clustering system 
is generated in Lagrangian space
\citep{1969ApJ...155..393P,1970Afz.....6..581D,1984ApJ...286...38W},
where the mass elements are described in their initial comoving coordinates.
It is also confirmed by many cosmological simulations that the tidal torque of 
protohalos (dark matter halos in Lagrangian space)
generated by the misalignment between the moment of inertia and the tidal field 
provides a persistent generation of angular momentum until virialization of halos
\citep{2002MNRAS.332..325P,2019PhRvD..99l3532Y}. 
Also, hydrodynamical simulations show that the spins of galaxies tend to align with 
their host halos \citep{2015ApJ...812...29T}.
These facts make galaxy spins another observable in constraining initial perturbations
\citep{2000ApJ...532L...5L,2001ApJ...555..106L}.
Reference \citep{2020PhRvL.124j1302Y} found a method to reconstruct galaxy spins
by initial perturbations, and the initial conditions can be estimated by
density reconstructions \citep{2014ApJ...794...94W}.
Reference \citep{2021NatAs...5..283M} applied this method and for the first time 
confirmed the correlation between galaxy spins and cosmic initial conditions.

Filaments are one of the largest 
structures of the cosmic web \citep{1982Natur.300..407Z,1996Natur.380..603B}.
By numerical simulations, they have been demonstrated to be spinning
in the LSS environment
\citep{2016MNRAS.460..816N,2020OJAp....3E...3N,2021MNRAS.506.1059X}.
Observationally, \citep{2021NatAs.tmp..114W} for the first time 
detected possible evidence for filament spins 
by examining the velocities of galaxies perpendicular to the filament’s axis.
These studies suggest that we could use the filament spins to understand the 
structure formation and potentially constrain cosmological models and parameters
using the framework similar to galaxies.
In Lagrangian space, the  
protofilaments could also be defined according to the mass elements
of filaments in their Lagrangian space.
Because filaments are generally much more massive than galaxies and halos,
they occupy larger regions in Lagrangian space, corresponding to larger,
more linear scales.
If there is a strong correlation between Eulerian and Lagrangian filaments,
they can be used to probe larger scales of the primordial perturbations
complimentary to that of galaxies and halos. 
It is thus interesting to examine the Lagrangian properties of filaments,
whether their initial spins can be described by the tidal torque theory,
and whether their spins are conserved across the cosmic evolution and then
can be reconstructed by the initial conditions.
In this paper, we use cosmological simulations to explore the Lagrangian properties
of cosmic filaments and their spin conservations.

The structure of the paper is as follows. 
In Sec.\ref{sec.simu}, we describe the simulation configurations and 
filament finder for filament identifications.
In Sec.\ref{sec.resu}, we present the results for the spin properties, 
conservations, and reconstructions.
In Sec.\ref{sec.conclu}, we give conclusions and make discussions and prospects. 

\section{Simulation and filament identification}\label{sec.simu}
We use numerical simulations to study the properties of filaments, 
which based on the cosmological $N$-body simulation code CUBE \citep{2018ApJS..237...24Y}. 
We assume a flat $\Lambda$CDM cosmology with cosmological parameters 
$\Omega_m=0.3$, $\Omega_{\Lambda}=0.7$, $\sigma_8=0.87$, $h=0.7$,
in a cubic box $L=100\,{\rm Mpc}/h$ per side with periodic boundary conditions.
$N_p=512^3$ particles are initially uniformly initialized in Lagrangian space, and the
``grid initial condition" is used where Lagrangian positions of particles
are placed at the each center of the cell, in a $N_g=512^3$ mesh,
so it is straightforward to acquire their Lagrangian properties.
The particles are then assigned with initial linear displacements and initial velocities
by using the Zel’dovich approximation \citep{1970A&A.....5...84Z} 
at initial redshift $z_{\rm init}=100$,
and then are evolved to Eulerian space at redshift $z=0$ 
using the particle-particle particle-mesh force calculation.
The particle mass is approximately $8.8\times10^8\Msun$.

Since filaments not self-bound by gravity and density dependency alone, 
there is no standard definition of cosmic filaments, 
resulting in many different filament finder realizations. 
According to their discrepancies of definitions, 
they can be classified by more mathematical and more physical ways. 
Mathematically, skeleton (e.g., the discrete persistent structure extractor 
\citep{2011MNRAS.414..350S,2011MNRAS.414..384S,2006MNRAS.366.1201N}), 
graph theory (e.g., state-of-the-art tracer T-Rex \citep{2020A&A...637A..18B}, 
MST \citep{2020MNRAS.499.4876P}), and 
Bayesian \citep{2014MNRAS.438.3465T,2007JRSSC..56....1S,2010A&A...510A..38S} 
methods extract the filament spine.  
Reference \citep{2018MNRAS.475.4494H} developed a novel method to find filaments in terms of 
machine learning. 
Many theories (pancake model, hierarchical clustering) interpret LSS formation
subjected to the tidal field tensor, 
defined by the Hessian matrix of the overdensity field,
$H_{ij}=\partial_i\partial_j\delta$,
where $\delta\equiv\rho/\langle\rho\rangle-1$ is the overdensity.
Accordingly many physical methods to trace cosmic web, 
including filaments, are based on Hessian matrix by
single scale \citep{2007MNRAS.375..489H,2010MNRAS.409..156B,2009MNRAS.396.1815F}
or multiscales \citep{2007A&A...474..315A,2013MNRAS.429.1286C,2014MNRAS.441.2923C}.

In this paper, we adopt the Smoothed Hessian Major Axis Filament Finder (SHMAFF)
\citep{2010MNRAS.409..156B}, which, physically, 
starts from the tensor field $H_{ij}$ to define the filament spine. 
Here we briefly introduce the primary parameters for completeness, 
and more details can be found in \citep{2010MNRAS.409..156B}.
For adapting the spine to the density field, we allocate all dark matter 
particles to a mesh with grid number $N_g=256^3$ by cloud-in-cell 
mass assignment, and smooth it with a $R_s = 2\,{\rm Mpc}/h$ Gaussian kernel. 
For each grid, we eigendecompose the matrix $H_{ij}$ 
into eigenvalues $\lambda_1<\lambda_2<\lambda_3$ and eigenvectors 
$\mathbf{A}_i\,(i=1,2,3)$. 
Intuitively, filament skeletons satisfy $\delta>0$ and $\lambda_2<0$, 
whereas $\mathbf{A}_3$ represents the alignment of the skeleton.\footnote{
  We note that $H_{ij}$ is parity even, so in
  the eigendecompositions, $\mathbf{A}_i$ is ``arrowless.''
}
We first remove all the grids satisfying any of the following criteria:
\begin{equation}\label{criteria.removal}
    \delta<0,\ \lambda_2 \geq 0.
\end{equation}
Then, we start from the grid at the minimum $\lambda_1$ and 
iteratively search for the adjacent grids along both directions, 
$\pm\mathbf{A_3}$, until the grid either satisfies 
the removal criteria (\ref{criteria.removal}),
or the $|\,\mathbf{A_3}\,|$ angle between two grid candidates 
exceeds a given threshold $C$, i.e.,
\begin{equation}
    |\,\mathbf{A}_{3,n}\times\mathbf{A}_{3,n-1}\,| > \sin(C\Delta),
\end{equation}
where $\Delta$ is the cell width. 
We take the angle value of $C = 30^\circ R_s^{-1}$. 
Note that we remove the grids of cylinder within width 
$W_i$ at each step as
\begin{equation}
    W_i = K\sqrt{\frac{-\rho_i}{\lambda_{1,i}}},
\end{equation}
where $K$ is set to 2 in this work.
The filament particles are identified by the criterion that
their vertical distance to the skeleton are less than
$2\,{\rm Mpc}/h$.

\begin{figure}[htb]
  \centering
   \includegraphics[width=0.95\linewidth]{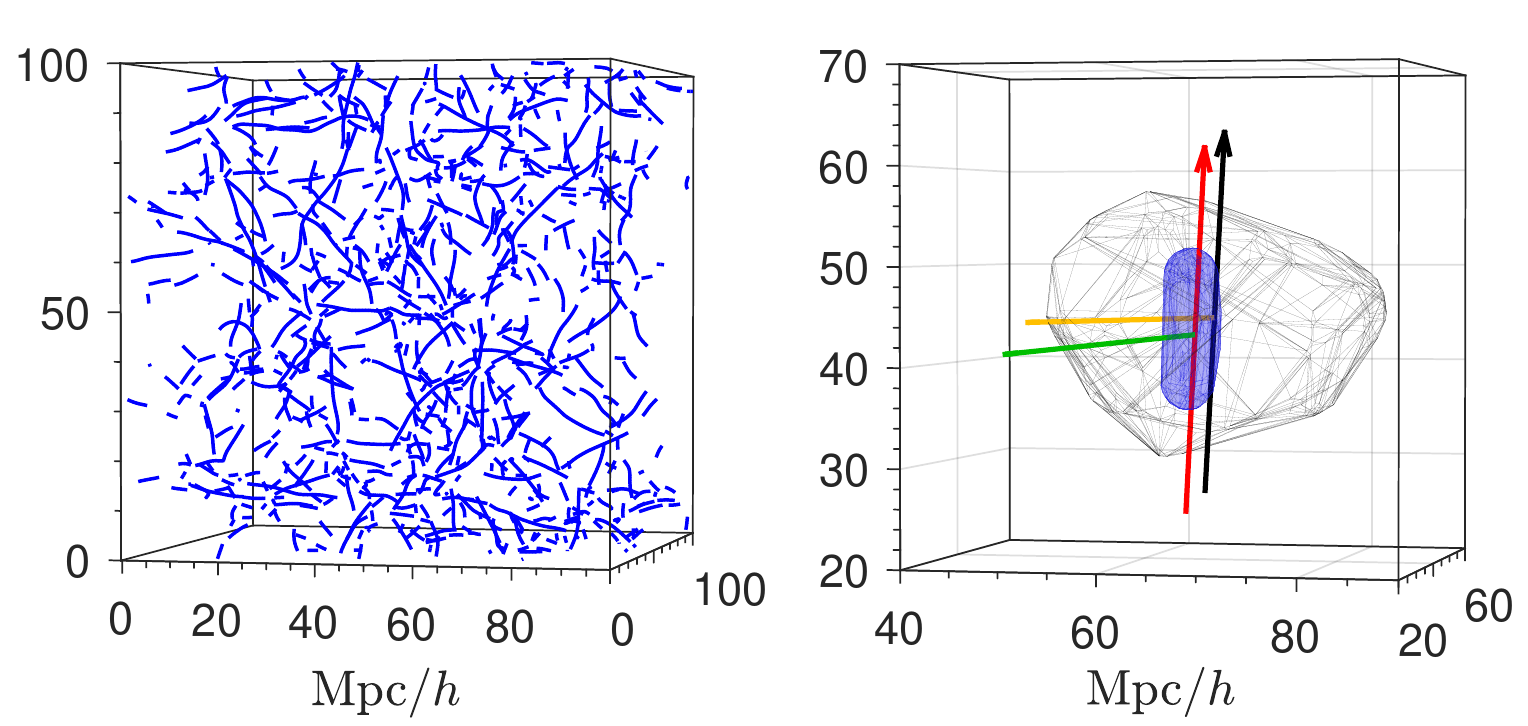}
   \caption{Visualization of filaments. 
   The left panel shows the skeletons of identified filaments in the simulation volume.
   The right panel shows the convex hull of a filament in Lagrangian (gray) 
   and Eulerian (blue) spaces.
   The mass of the selected filament is $1.05\times10^{15}\Msun$.
   The black and red arrows represent the direction of spines 
   of the filament in Lagrangian and Eulerian spaces and the yellow and green
   arrows represent the directions of their angular momenta, respectively. }
   \label{fig.filament}
\end{figure}

By applying the filament finder we get a filament catalog with 1680 filaments. 
This catalog contains the detailed information about the filament skeletons 
and particles. In the left panel of Fig.\ref{fig.filament}, 
we plot the skeletons of all filaments in the simulation box. 

\section{results}\label{sec.resu}

\subsection{Mass distributions}

In this section we start with a comparison between filament properties
in Eulerian and Lagrangian spaces.
The protofilaments are obtained by using particle IDs in the 
$N$-body simulation and tracing back to their Lagrangian positions.
We use moment of inertia tensor
$I_{ij}=\sum_i m_i {x}'_i{x}'_j$
to characterize the mass distribution of a system up to quadrupole.
Here $m_i$ is the particle mass, and
${\bs x}'$ is the particle position relative to
the center of mass in Eulerian or Lagrangian space in consideration.
The eigendecomposition of $I_{ij}$ gives the the primary, intermediate, and minor 
axes of the mass distribution of the filament and their alignments in space.
The eigenvalues are sorted as $i_1>i_2>i_3$, so the the primary axis $i_1$,
associated with the eigenvector ${\bold V}_1$
(hereafter we denote $\bold V_j$ as the eigenvector of $i_j$, $j=1,2,3$)
is expected to align with the spine of the filament. 
The properties of the mass distribution can be characterized by three parameters, 
the trace $\tau=i_1+i_2+i_3$, the ellipticity $e=(i_1-i_3)/2\tau$, 
and the prolateness $p=(i_1-2i_2+i_3)/2\tau$ \citep{2002MNRAS.332..339P}. 
A perfect sphere has $e=p=0$,
a thin disk has $e=1/4$ and $p=-1/4$,
while a slim straight filament has $e=1/2$ and $p=1/2$. 

\begin{figure}[htb]
  \centering
  \includegraphics[width=1.0\linewidth]{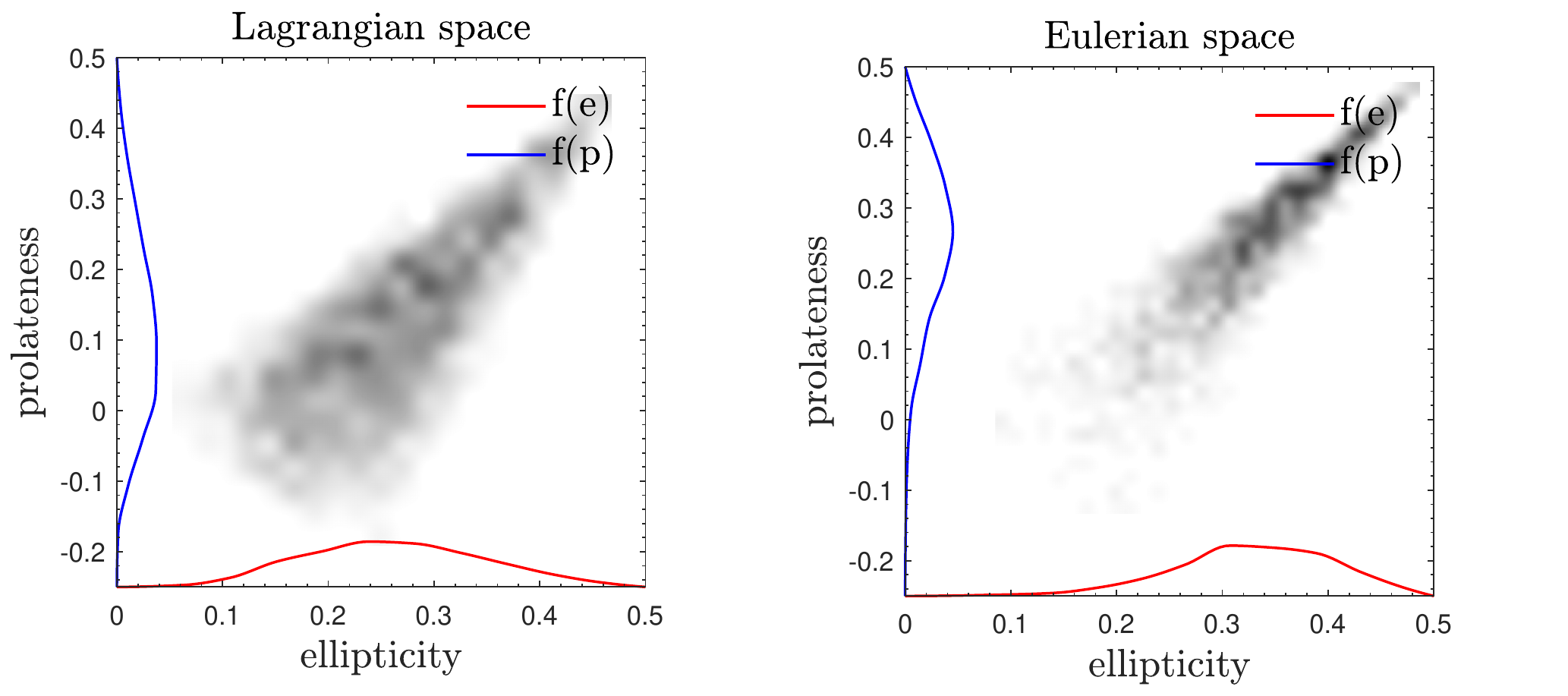}
  \includegraphics[width=1.0\linewidth]{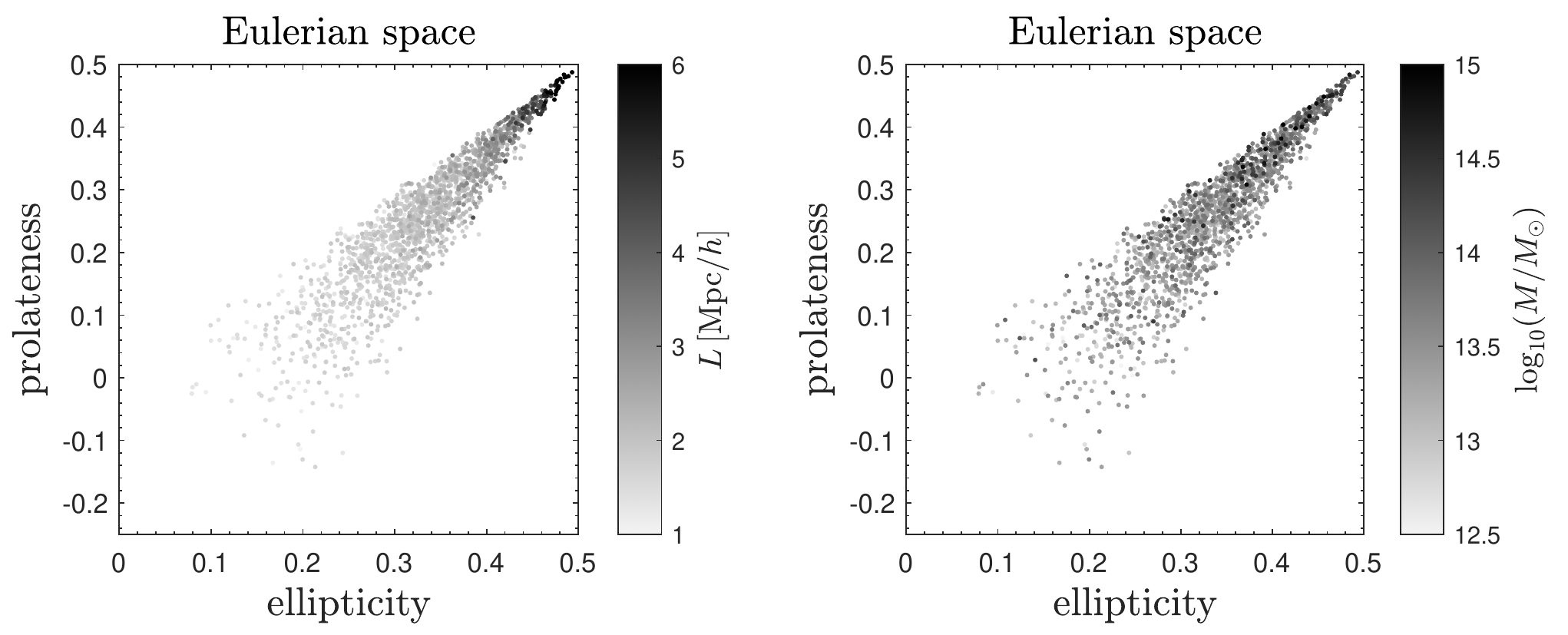}
   \caption{Joint PDFs of 
   ellipticity $e$ and prolateness $p$ for filaments 
   in Lagrangian (top left panel) and Eulerian (top right panel) spaces,
   shown by gray scale. 
   The red and blue curves represent the PDFs 
   of $e$ and $p$, respectively.
   The bottom panels are the same as the top right panel, but points are colored according 
   to the spine length of the filaments (bottom left panel) and the mass of the filaments 
   (bottom right panel).}
   \label{fig.ep}
\end{figure}

In the top panels of Fig.\ref{fig.ep}, we plot 
the joint probability distribution functions (PDFs) of 
ellipticity $e$ and prolateness $p$ for all 1680 filaments 
in Lagrangian and Eulerian spaces, respectively, 
and the one-dimensional PDF marginalized along each axis. 
As expected, in Eulerian space, nearly all filaments show prolate ($p>0$)
mass distributions rather than oblate ($p<0$).
In terms of the expectation value of the distribution,
$\left\langle p_{\rm Eul}\right\rangle=0.25$.
Meanwhile, the filaments show systematic ellipticity, 
$\left\langle e_{\rm Eul}\right\rangle=0.33$.
In the bottom panels of Fig.\ref{fig.ep}, we plot the dependence of 
$e$ and $p$ of filaments in Eulerian space
with spine length and mass respectively, where spine length is represented
by $\sqrt{i_1}$.
Filaments with longer spine length tend to have 
$(e,p)\rightarrow (1/2,1/2)$, but this trend shows weak dependence on mass. 
Note that there are still few filaments with low or negative prolateness,
and with low ellipticity.
These filaments are generally shorter and less massive and the inclusion of 
particles involve numerical artifacts.
However in later subsections they do not affect the conclusions.
In contrast, the protofilaments in Lagrangian space tend to be more 
spherical relatively.
In the top left panel of Fig.\ref{fig.ep}, the joint PDF does not cluster
to the $(e,p)=(1/2,1/2)$ corner.
Numerically the statistics of these two parameters in Lagrangian space are
$\left\langle p_{\rm Lag}\right\rangle=0.12$ and 
$\left\langle e_{\rm Lag}\right\rangle=0.26$.

These results illustrate the filament formation picture in the structure evolution. 
The protofilament region is more spherical rather than filamentary initially. 
While in contrast to protohalos, which collapse in all directions,
protofilaments primarily collapse along two directions, $\bold V_2$ and $\bold V_3$,
due to the external tidal field.
Because the initial spin given by the tidal torque is preferentially
aligned with $\bold V_2$ and the second principal axis of the 
tidal tensor $\bold T_2$ \citep{2001ApJ...555..106L}
(hereafter we denote $\bold T_i$ and $t_i$ as the eigenvectors
and eigenvalues of $\bold T$),
we thus expect that in the anisotropic collapse of filaments,
the tidal torque is also preferentially aligned with $\bold V_2$.
In the following section we indeed find that both Lagrangian and Eulerian filament spins 
are preferentially aligned with $\bold V_2$; i.e., the spin vectors are preferentially
perpendicular to the spine of the filaments $\bold V_1$.
Besides, a spherical Lagrangian region is potentially suitable to directly apply 
the spin reconstruction methods presented in \citep{2020PhRvL.124j1302Y}.

\subsection{Spin directions and conservations}
Here we study the angular momentum properties of filaments.
We use $\bs j$ to denote the angular momentum vector.
In Eulerian space, the angular momentum vector of 
a filament is defined as 
$\bs j_E=\sum_i m_i (\bs x_i - \overline{\bs x})\times \bs v_i=
\sum_i m_i\bs x'_i \times \bs v_i$,
where $m_i$, $\bs x_i$, $\overline{\bs x}$ and $\bs v_i$ are 
the particle mass, Eulerian position, Eulerian center of mass, 
and Eulerian velocity, respectively. 
$\bs x'_i \equiv \bs x_i - \overline{\bs x}$ is the position relative to
the center of mass $\overline{\bs x}$. 
In Lagrangian space, the angular momentum vector is similarly defined as 
$\bs j_L=\sum_i {m_i \bs q'_i \times \bs u_i}
=\sum_i {m_i (\bs q_i - \overline{\bs q}) \times \bs u_i}
=\sum_i {m_i \bs q'_i \times (-\nabla \phi |_{\bs q_i})}$, 
where $\bs q_i$, $\overline{\bs q}$, $\bs u_i$ are the Lagrangian position, 
Lagrangian center of mass, Lagrangian velocity, respectively.
$\bs q'_i\equiv \bs q_i - \overline{\bs q}$ is the Lagrangian position
relative to Lagrangian center of mass $\overline{\bs q}$.
The Lagrangian velocity $\bs u_i$ is simply expressed by 
the gradient of the primordial gravitational potential $\phi$,
consistent with our setup in the initial conditions \citep{1970A&A.....5...84Z}. 

\begin{figure}[htb]
  \centering
   \includegraphics[width=0.95\linewidth]{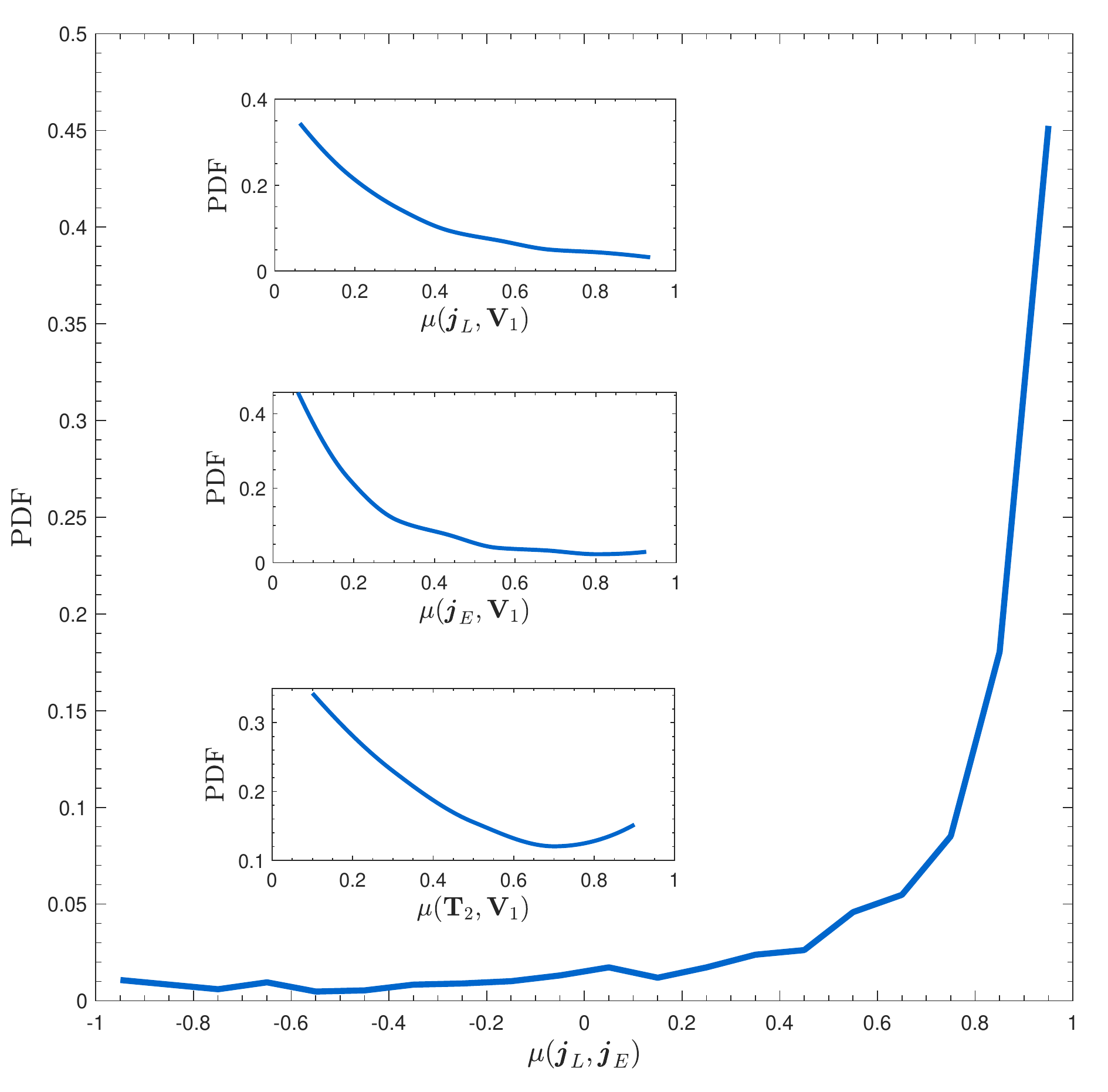}
   \caption{Spin conservation of cosmic filaments shown by PDF of $\mu(j_L,j_E)$. 
   The distribution shows a strong deviation from a top-hat distribution, 
   suggesting that the filament spin directions are conserved from high redshifts
   to low redshifts.
   The three insets are the PDFs of $\mu(j_L,\bold V_1)$, 
   $\mu(j_E,\bold V_1)$ and $\mu(\bold T_2,\bold V_1)$, where $\bold T_2$ denotes the 
   second principal axis of the tidal tensor.}
   \label{fig.cc}
\end{figure}

We use the cosine of the angle between two vectors $\bs j_L$ and $\bs j_E$ 
to quantify the cross-correlation between their directions, i.e., 
$\mu(\bs j_L,\bs j_E) \equiv \bs j_L \cdot \bs j_E / |\,\bs j_L\,|\,|\,\bs j_E\,| 
\in [-1,1]$.
The $\mu$ of two randomly distributed vectors in three-dimensional
space is top-hat distributed between $-1$ and $1$, with expectation $0$.
In Fig.\ref{fig.cc}, we plot the PDF of $\mu(\bs j_L,\bs j_E)$ for all filaments. 
The expectation value $\langle \mu \rangle $ takes 0.70 and the PDF of 
$\mu(\bs j_L,\bs j_E)$ obviously depart from a top-hat distribution, 
suggesting that $\bs j_L,\bs j_E$ directions are strongly correlated.
This property is similar to the spin conservations of dark matter halos
\citep{2021PhRvD.103f3522W}.
Next we decompose the spin vector into parallel and perpendicular components
with respect to the major axis ${\bold V}_1$ (spine) of the Eulerian filament.
From now on, we denote them with superscripts $\parallel$ and $\perp$.
In the first two insets of Fig.\ref{fig.cc}, we plot the PDFs of
$|\,\bs j_L^{\,\parallel}\,|/|\,\bs j_L\,|=|\,\mu(\bs j_L,\bold V_1)\,|$
and $|\,\bs j_E^{\,\parallel}\,|/|\,\bs j_E\,|=|\,\mu(\bs j_E,\bold V_1)\,|$,
which show that in both Lagrangian and Eulerian spaces,
the spin directions are preferably perpendicular to the spine
of the Eulerian filaments. 
This is consistently explained by the tidal torque theory,
which expresses the initial tidal torque spin as
$j_{i}\propto\epsilon_{ijk}I_{jl}T_{lk}$, where $\epsilon_{ijk}$
is the Levi-Civita symbol and $T_{lk}\propto-\partial_l\partial_k\phi$
is the tidal tensor.
In the coordinate system of principal axes of $\bold T$,
$j_2\propto(t_1-t_3)I_{31}$ is the dominated component due to the 
largeness of $t_1-t_3$ [for more details, see Eq.(2) and the following 
discussions of \citep{2001ApJ...555..106L}].
In comparison, the filament spine $\bold V_1$ is aligned with the 
least collapsing direction $\bold T_1$, and $\bold T_1 \perp \bold T_2$.
The third inset of Fig.\ref{fig.cc} confirms it numerically by 
the PDF of $\mu(\bold T_2,\bold V_1)$ and thus 
explains the dominated spin component $j_2\perp\bold V_1$ statistically.

To understand the spin directions more intuitively, 
we select a typical filament with mass $1.05\times10^{15}\Msun$ 
in our simulation and visualize
its shapes in Lagrangian and Eulerian spaces,
as well as the spin vectors in both spaces 
in the right panel of Fig.\ref{fig.filament}.
The shape is visualized by the convex hull of all particles
belonging to the filament, either in Eulerian (blue) or Lagrangian (gray) space.
Consistent with the statistics of Fig.\ref{fig.ep},
the Eulerian filament is elongated vertically, whereas its
Lagrangian counterpart is more spherical.
The yellow and green arrows show their spins, and
the black and red arrows represent their components in the major axis ${\bold V}_1$.
Their magnitudes are normalized arbitrarily for better visualization.
The conservation of spin directions is seen by this typical filament.

\subsection{Spin magnitudes}

In this subsection we focus on the conservation of spin magnitudes.
The comparison of angular momentum magnitudes through the cosmic
evolution is conveniently analyzed by the dimensionless spin parameter.
The spin magnitude of a system in Eulerian space can be characterized by 
a dimensionless kinematic spin parameter $\lambda_{KE}$ \citep{2021PhRvD.103f3522W}, 
which is defined as 
\begin{equation}\label{eq.lambdaK}
  \lambda_{KE} \equiv \frac{\int_{V_x} \hat{j}_i \epsilon_{ijk} x'_j v'_k \,{\rm d}M}{\int_{V_x} x' v' \,{\rm d}M}=
  \frac{\int_{V_x} \sin\theta_1 \cos\theta_2 x' v' \,{\rm d}M}{\int_{V_x} x' v' \,{\rm d}M},    
\end{equation}
where $\hat{j}=(\bs j_E/j_E)$ is the unit $\bs j_E$ vector $\left(j_E=|\,\bs j_E\,|\right) $, $x'=|\bs x'|$,
 $v'=|\bs v'| \equiv  |{\bs v}-\overline{\bs v}|$ is the velocity relative to the 
average velocity of the filament
 $\overline{\bs v}$, $\sin\theta_1=\sqrt{1-\mu^2{(\bs x',\bs v')}}$ and $\cos\theta_2=
\mu\left(\bs x' \times \bs v',\bs j_E\right)$, which can be similarly defined 
in Lagrangian space and denoted with $\lambda_{KL}$. 
They take the value $[0,1]$ and describe whether a system is more
velocity dispersion supported or rotation supported. 
For dark matter halos and their protohalos in Lagrangian space,
\citep{2021PhRvD.103f3522W} demonstrated that both spin directions
and spin magnitudes tend to be correlated across cosmic evolution.

For two variables $X,Y$, the correlation coefficient is defined as 
\begin{equation}\label{eq.r}
  r(X,Y)
  =\frac{{\rm Cov}(X,Y)}
  {\sqrt{{\rm Var}[X]\cdot{\rm Var}[Y]}},
\end{equation}
and the covariance is normalized by the square root of the multiplication
of their autovariances.
The correlation $r\in[-1,1]$, statistically
$r=\pm 1$, indicates the strongest correlation/anticorrelation,
and $r\simeq 0$ indicates a noncorrelation.

\begin{figure}[t]
  \centering
   \includegraphics[width=0.95\linewidth]{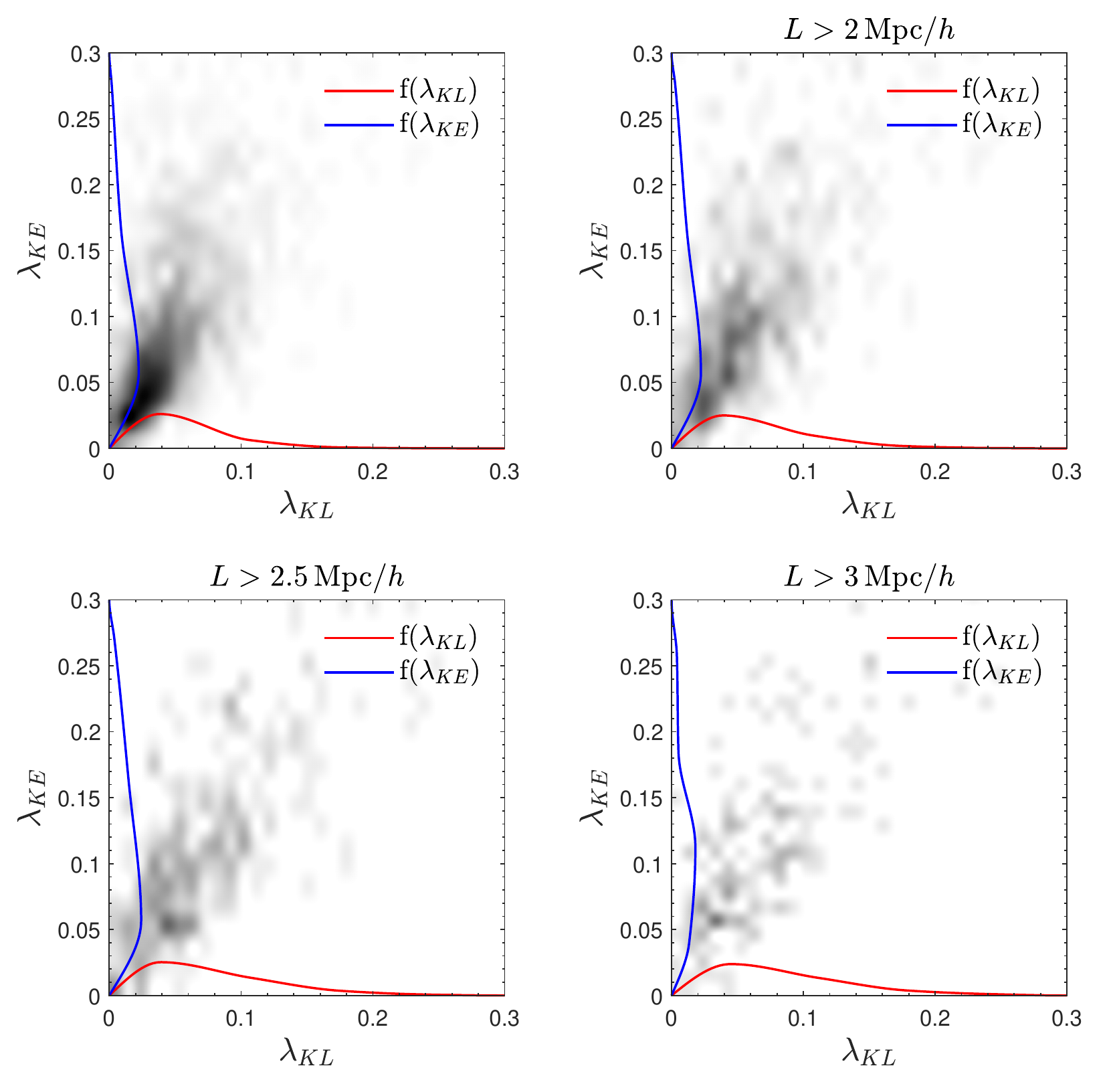}
   \caption{Correlation between $\lambda_{KL}$ and $\lambda_{KE}$ 
   for filaments in different length ranges.
   The gray scale in the background represents their joint distribution and the red and blue curves 
   represent the PDFs of the parameters indicated by the label of each axis.}
   \label{fig.lambdaK}
\end{figure}

In Fig.\ref{fig.lambdaK}, we plot the correlation between 
$\lambda_{KL}$ and $\lambda_{KE}$ for filaments in different length ranges. 
Table \ref{Tab1} lists the number of filament samples and the correlation coefficient
$r(\lambda_{KL},\lambda_{KL})$ as a function of filament spine length range.
Figure \ref{fig.lambdaK} and Table \ref{Tab1} suggest that the spin magnitudes of
$\bs j_{L}$ and $\bs j_{E}$ have a strong positive correlation.  
We find that this positive correlation is related to the length of the spine. 
Filaments with longer spines tend to have a stronger 
correlation between $\lambda_{KL}$ and $\lambda_{KE}$.
This trend can be explained by the fact that longer filaments tend to be more filamentary in our 
samples as we have shown in the bottom left panel of Fig.\ref{fig.ep},
which is partly a consequence of the limitation of the filament finder. 
By eliminating those filaments with low prolateness and ellipticity could make the 
correlation more robust.

By observing the PDFs of single parameters, we find that
the values of 
$\lambda_{L}$ and 
$\lambda_{E}$ are much less than unity,
suggesting that the filaments are not rotation supported in both 
Eulerian and Lagrangian spaces.
This behavior is very similar to dark matter halos and protohalos 
by various of definitions \citep{2021PhRvD.103f3522W}.

\begin{table*}[t]
  \centering
  \caption{Number of filament samples and the correlation coefficient as a function of 
  filament spine length range.
    \label{Tab1}}
  \renewcommand\arraystretch{2}
  \centering
  \begin{tabular}{m{2.5cm}<{\centering}m{2cm}<{\centering}m{2cm}<{\centering}m{2cm}<{\centering}m{2cm}<{\centering}}
  \hline
  {Spine length} & {$L>0\,{\rm Mpc}/h$} & {$L>2\,{\rm Mpc}/h$} & {$L>2.5\,{\rm Mpc}/h$} & {$L>3\,{\rm Mpc}/h$} \\
  \hline
  {Sample number} & 1680 & 746 & 326 & 177\\
  {$r(\lambda_{KL},\lambda_{KE})$} & 0.57 & 0.61 & 0.712 & 0.717\\
  {$r(\lambda_{L}^{\,\parallel},\lambda_{E}^{\,\parallel})$} & 0.67 & 0.70 & 0.771 & 0.774\\
  \hline
  \end{tabular}
\end{table*}

To compare with the current observation works 
\citep{2021NatAs.tmp..114W}, which detected observational evidence for 
$\bs j^{\,\parallel}$, it is interesting to extract the spin magnitudes of 
$\bs j^{\,\parallel}$, which is parallel to the spine (${\bold V}_1$), 
and check whether it is correlated to their spin magnitudes in Lagrangian space.
By projecting Eq.(\ref{eq.lambdaK}) onto a plane perpendicular to ${\bold V}_1$,
the kinematic spin parameter becomes 
\begin{equation}\label{eq.lambdaKE}
  \lambda_{E}^{\,\parallel} \equiv \frac{ \hat{\bold V}_1\cdot \bs j_E^{\,\parallel} } 
  {\sum{ x'^\perp v'^\perp}},
\end{equation}
where $x'^\perp=|\bs x'^\perp|$,
$v'^\perp=|\bs v'^\perp|$, vectors with $_\perp$ are the components
perpendicular to the spine.
Here $\hat{\bold V}_1$ is an unit vector aligned with the spine.
Because the spine is parity even, the sign of $\lambda_{E}^{\,\parallel}$
is defined according to the arbitrarily chosen  $\hat{\bold V}_1$.
Similarly, $\lambda_{L}^{\,\parallel}$ can be obtained from $\bs j_L^{\,\parallel}$ 
in the same manner,
\begin{equation}\label{eq.lambdaKL}
  \lambda_{L}^{\,\parallel} \equiv \frac{ \hat{\bold V}_1\cdot \bs j_L^{\,\parallel} } 
  {\sum{ q'^\perp u'^\perp}}.
\end{equation}
With the above definition, $|\,\lambda_{L}^{\,\parallel}\,|,|\,\lambda_{E}^{\,\parallel}\,|
\in [0,1]$, and larger $|\,\lambda_{L}^{\,\parallel}\,|$ or $|\,\lambda_{E}^{\,\parallel}\,|$
corresponds to more coherent rotations along the spine.
Meanwhile, same/different signs of $\lambda_{L}^{\,\parallel}$ 
and $\lambda_{E}^{\,\parallel}$ indicates that $\bs j_E^{\,\parallel}$ is 
parallel/antiparallel to $\bs j_L^{\,\parallel}$.

\begin{figure}[t]
  \centering
   \includegraphics[width=0.95\linewidth]{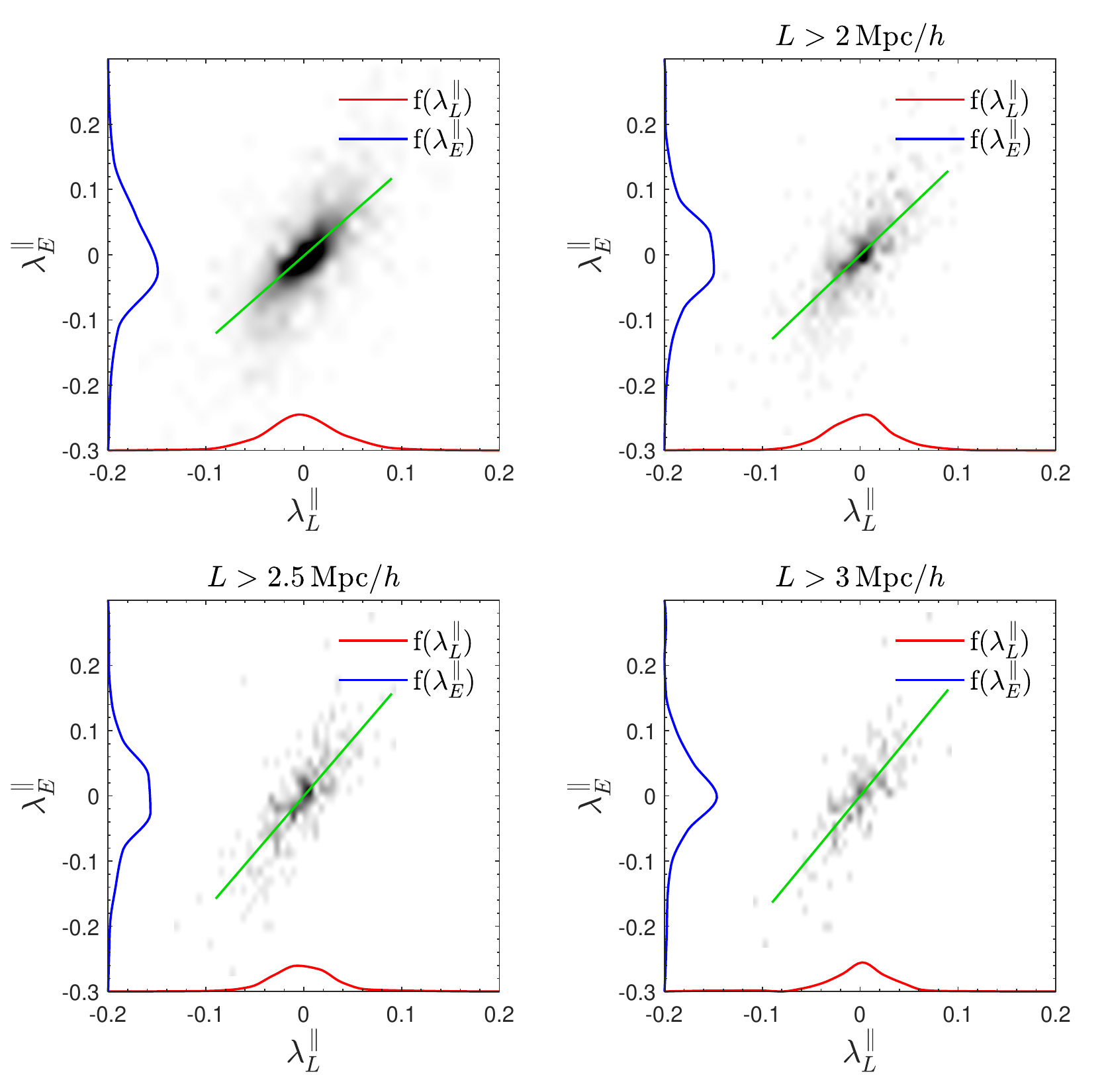}
   \caption{The same as Fig.\ref{fig.lambdaK} but for $\lambda_{L}^{\,\parallel}$ and 
   $\lambda_{E}^{\,\parallel}$. The green lines are the linear fittings of all data points.}
   \label{fig.lambdaP}
\end{figure}

In Fig.\ref{fig.lambdaP}, we plot the correlation between 
$\lambda_{L}^{\,\parallel}$ and $\lambda_{E}^{\,\parallel}$ for filaments in 
different length ranges. The correlation coefficients 
$r(\lambda_{L}^{\,\parallel},\lambda_{E}^{\,\parallel})$ in different spine 
length range are also listed in Table \ref{Tab1}. We find that the parallel 
component of the spin shows similar correlation as $\bs j$, the spin magnitudes of
$\bs j_{L}^{\,\parallel}$ and $\bs j_{E}^{\,\parallel}$ have a strong positive
correlation and filaments with longer spines tend to have a stronger correlation.
Besides, the absolute values of $\lambda_{L}^{\,\parallel}$ and 
$\lambda_{E}^{\,\parallel}$ are also much less than unity, but fortunately 
the smallness of the rotation component of the filaments can still be measured
\citep{2021NatAs.tmp..114W}.

\subsection{Spin reconstruction}

\begin{figure}[htb]
  \centering
   \includegraphics[width=0.95\linewidth]{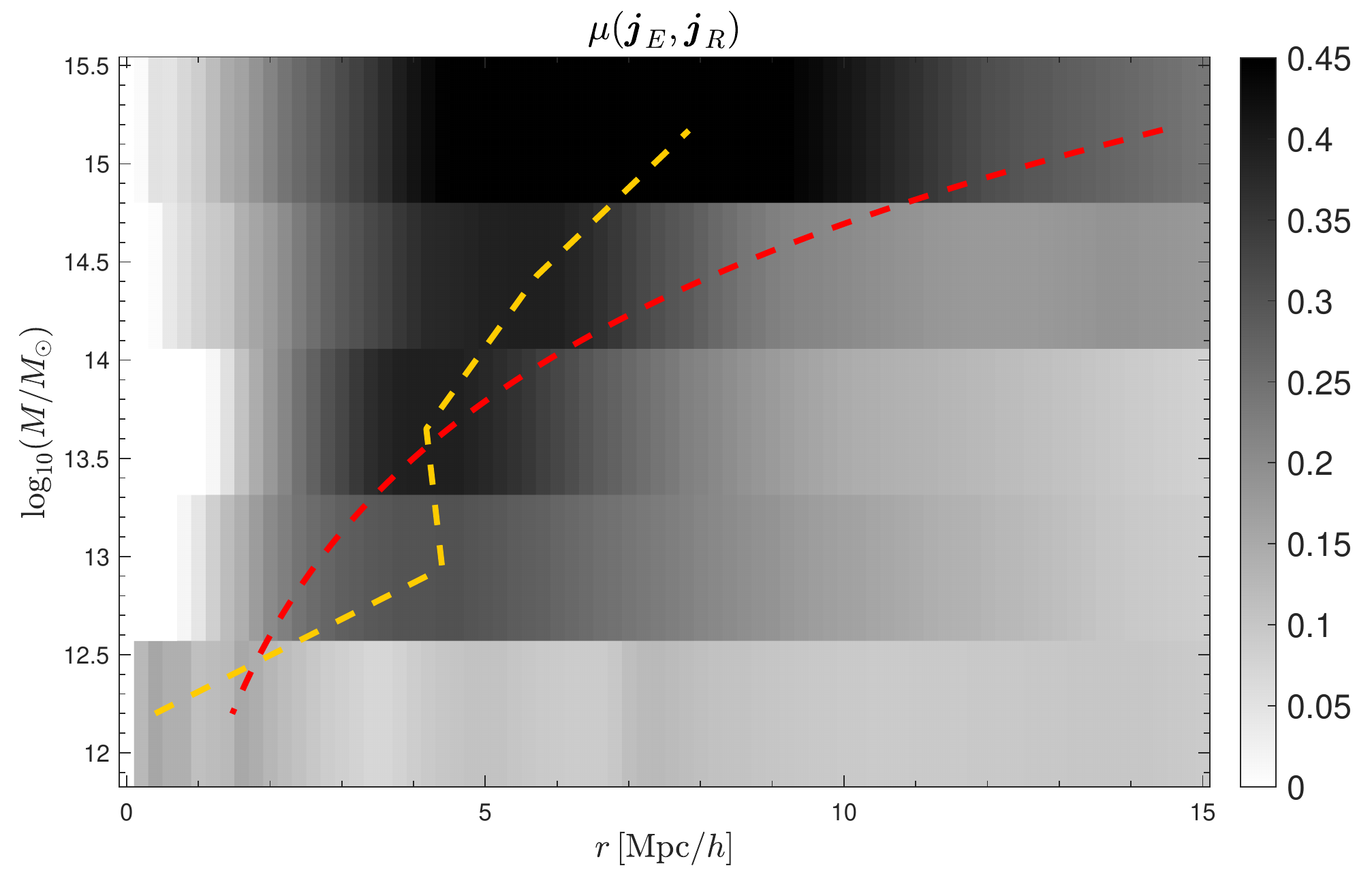}
   \caption{Correlation between the Eulerian $\bs j_E$ filament spins and spins reconstructed 
   using Eq.(\ref{eq.jR}) from known initial conditions $\bs j_{R}$. 
   The gray scale in the background shows the degree of correlation and darker colors show better reconstruction.
   The red and yellow dashed represent the Lagrangian equivalent protofilament radius $r_q$ and 
   optimal smoothing scale $r_{\rm opt}$, respectively.}
   \label{fig.corr_x}
\end{figure}

In this subsection we reconstruct the predicted spins for 
filaments based on their Lagrangian space properties
analogous to the the spin reconstruction of halos.
As we have mentioned in Sec.\ref{sec.resu},
the initial angular momentum vector of a protohalo that 
initially occupies Lagrangian volume $V_L$ is approximately by 
$\bs j \propto \epsilon_{ijk}I_{jl}T_{lk}$, 
where $\mathbf{I}=(I_{jl})$ is the moment of inertia tensor of $V_L$, 
$\mathbf{T}=(T_{lk})$ is the tidal tensor acting on $\mathbf{I}$, 
and $\epsilon_{ijk}$ is the Levi-Civita symbol collecting the antisymmetric 
components generated by the misalignment between $\mathbf{I}$ and $\mathbf{T}$. 
The spin reconstruction of dark matter halos is by defining \citep{2020PhRvL.124j1302Y}
\begin{equation}\label{eq.jR}
  \bs j_R=(j_i)\propto\epsilon_{ijk}\bs \T_{jl}\bs \T^{+}_{lk},
\end{equation}
where $\bs\T$,$\bs \T^{+}$ are tidal fields constructed as Hessians of the 
initial gravitational potential smoothed at two different scales $r,R$. 
The initial gravitational potential can be estimated by the initial density field reconstructed 
method ELUCID, for which we refer the readers to \citep{2014ApJ...794...94W,2016ApJ...831..164W} 
for more details. This method can reproduce the density field of the
nearby universe generated from the galaxy distributions in observation.
To obtain $\bs \T$ and $\bs \T^{+}$, we smooth the initial gravitational potential $\phi_{\rm init}
(\bs q)$, by multiplying it in the Fourier space by the baryonic acoustic oscillation damping model 
${\mathcal D}(k)^{1/4}=\exp{(-r^2k^{2}/2)^{1/4}}$  \citep{2017PhRvD..95d3501Y}. 
By choosing $R \to r_{+}$, we find $\bs j_R$ a good approximation for an angular 
momentum of a protohalo. 
Similarly, we apply this method to protofilaments, 
then we get the spin field $\bs j_{R}$ reconstructed from known initial conditions.
We group all identified filaments into five mass bins, ranging from
$\sim 10^{12}\Msun$ to $\sim 10^{15.5}\Msun$, then apply 
Eq.(\ref{eq.jR}) with a set of different smoothing scales $r$,
and compute the correlations between reconstructed and Eulerian spins of filaments.
In Fig.\ref{fig.corr_x}, we plot $\mu(\bs j_E,\bs j_R)$
as a function of smoothing scales $r$ and filament mass bins.
The gray scale shows the degree of correlation and the darkest region in 
each mass bin represents the optimal smoothing scale $r_{\rm opt}$,
also indicated by the yellow dashed curve.

We find that the spins of more massive filaments can generally be
predicted at a wide range of smoothing scales.
The correlation is larger than 0.3 for filaments more
massive than $10^{13.5}\Msun$.
We also find that more massive filaments are better reconstructed 
by a larger smoothing radius, and this behavior is similar to
the spin reconstruction of dark matter halos \citep{2020PhRvL.124j1302Y}.
As a reference, we plot with the red dashed curve the equivalent protofilament radius 
in the Lagrangian space defined as $r_q\equiv(2MG/\Omega_{m}H^{2}_{0})^{1/3}$. 
We find that $r_{\rm opt}$ is not closely related to $r_q$,
and find $\sim 5\,{\rm Mpc}/h$ a universal smoothing scale for
all massive filaments. 

\section{conclusion and discussions}\label{sec.conclu}
  In this paper, by using numerical simulations, we study the
  angular momentum properties of cosmic filaments across the cosmic
  evolution, as well as their origins, conservations, and predictability.
  The conclusion of our results is summarized as follows:
  \begin{itemize}
    \item In terms of moment of inertia tensors and their eigendecompositions,
          the cosmic filaments in Lagrangian space (protofilaments) exhibit
          much more spherical shapes, for which Lagrangian spin reconstruction
          method with a isotropic smoothing function is effective to be applied
          for the angular momentum prediction.
    \item The angular momentum directions of filaments and their protofilaments
          are very well correlated, with a statistical correlation of $0.7$,
          and significantly depart from a random distribution from
          uncorrelated pairs of vectors. This shows that the angular momentum
          directions of filaments are well conserved through the cosmic
          evolution, similar to that of dark matter halos.
    \item The angular momentum direction is more perpendicular to the 
          spine (major axis) of the filament, which can be well predicted by 
          tidal torque theory, whereas the spin component parallel to 
          the spine matches the numerical and observational analysis of 
          filaments in previous studies.
    \item By constructing a dimensionless spin parameter of this spin and its parallel 
          component, we find that the kinematic motion
          of the filaments relatively less rotation supported. This statistics
          is very similar to that of dark matter halos.
    \item The dimensionless spin magnitudes of protofilaments and filaments are
          statistically significantly correlated, showing that faster spinning
          protofilaments are more likely to form faster spinning filaments at
          low redshifts.
    \item The filament spins can be predicted by a spin reconstruction method 
          in Lagrangian space, and the predictability is similar to the spin
          reconstruction of dark matter halos. This opens up the possibility
          of using filament spins to constrain the cosmic initial conditions.
  \end{itemize}

  We notice that the above conclusions weakly depend on the
  mass and length of the filaments, with longer and more massive filaments having
  better spin conservation and predictability. This can be partly explained
  by the limitation of filament finders, meaning that those short and low massive 
  samples in the catalog might be fake filament structures and that eliminating them could 
  make the results more robust.
  The free parameters in our filament finder and other available filament
  finder algorithms add freedom in the identifications of filaments and
  their containing particles.
  As a convergence test, we also use different parameters in our filament
  finder discussed in Sec.\ref{sec.simu} and they all give consistent
  results. Discovering other filament finder methods and comparing the results
  are not included in this paper and can be left to future studies.
  
  However, it is more important to study how the filament spins can be observed 
  by multiple tracers, such as galaxies and their relative velocities,
  e.g., \citep{2021NatAs.tmp..114W},
  or intergalactic media by kinetic Sunyaev Zel’dovich effect \citep{2019JCAP...06..001B}.
  Galaxy formation simulations in a cosmological volume are needed to understand
  these effects and are helpful to construct the pipeline of the analysis.
  Moreover, besides \citep{2021NatAs.tmp..114W,2021MNRAS.506.1059X},
  it is also valuable to extract the angular momenta perpendicular to the
  spines of the filaments.
  Multitracer reconstruction of the initial density field and 
  redshift space distortion should be included in the analysis.
  We leave them to future works.

\section*{ACKNOWLEDGMENTS}
We thank the anonymous referee for valuable suggestions.
This work is supported by National Science Foundation of China Grants No. 11903021 and 
No. 12173030.
P.W. and  X.K. acknowledge support from the joint Sino-German DFG 
research Project
DFG-LI 2015/5-1, NSFC No. 11861131006.
The simulations were performed on the workstation of cosmological sciences,
Department of Astronomy, Xiamen University.

\bibliographystyle{h-physrev3}
\bibliography{mingjie_ref}

\end{document}